Low Temperature Thermoelectric Properties of Co- and Cr- doped CuAgSe

Peter Czajka, Mengliang Yao, Cyril Opeil
Department of Physics, Boston College

**Abstract**

High mobility phonon-glass semimetal CuAgSe has shown promise in recent years as a potential low-temperature thermoelectric material. It exhibits reasonably strong thermoelectric performance as well as an extremely high carrier mobility, both of which are enhanced when the material is doped with Ni at the Cu sites. The exact mechanism by which these enhancements result; however, is unclear. In order to further investigate the effects of chemical substitution on the material's thermoelectric properties, we have prepared and performed various measurements on CuAgSe samples doped with Co and Cr according to the following compositional formulas: $Cu_{1-x}Co_xAgSe$ (x=0.02, 0.05, 0.10) and $Cu_{1-x}Cr_xAgSe$ (x=0.02, 0.05). Measurements of temperature and magnetic field dependent thermal conductivity, electrical resistivity, and Seebeck coefficient will be discussed. Our results reveal a remarkable sensitivity of CuAgSe's thermoelectric properties to chemical doping in general as well as a particular sensitivity to specific dopants. This demonstrated tunability of CuAgSe's various properties furthers the case that high mobility phonon glass-semimetals are strong candidates for potential low temperature thermoelectric applications.

*Keywords:* CuAgSe, thermoelectric, semimetal, low temperature.

**1. Introduction**

The last three decades have been incredibly prolific for the field of thermoelectrics [1-4]. Thermoelectric efficiency is typically quantified using a dimensionless figure of merit called ZT, the formula for which is:

$$ZT = \frac{S^2 \sigma}{\kappa} T$$

where S is the Seebeck coefficient ($\Delta V/\Delta T$), $\sigma$ is the electrical conductivity, $\kappa$ is the thermal conductivity, and T is the absolute temperature. This means that a good thermoelectric material must conduct electricity well and heat poorly in addition to exhibiting the Seebeck to a significant degree. However, while ZT values at or near the commercial viability threshold value of 2.5 have recently begun to be achieved with high temperature materials such as SnSe [5] and PbTe [6], efficiencies at low temperatures are still relatively poor. High efficiencies around and below 77 K (the boiling point of Nitrogen) are particularly desirable because they could allow thermoelectrics to be used for cryogen-free cryogenic solid-state cooling [7-8]. Most of the materials that have been studied in context of cryogenic thermoelectric applications are either conventional sorts of semiconductors such as Bi-Sb alloys [9-10] and $CsBi_4Te_6$ [11] or rare-earth element containing systems that exhibit significant correlated electron physics such as $CePd_3$ [12], $CeAl_3$ [13], and $YbAgCu_4$ [14]. Until recently, semimetals had generally been ignored in thermoelectrics research because the dual nature of their carrier populations usually leads to poor thermoelectric performance. However, several promising candidates have recently been either identified or proposed. These include various Zintl phase [15-16] and half-Heusler [17-18] materials as well as several of the Dirac and Weyl type semimetals [19-21] inspired by the recent "topological revolution" in condensed matter physics. Because semimetals have been studied so little in this context (apart from those with significant correlated electron effects), the nature of thermoelectric phenomena in these sorts of systems is not well-understood.

One such material that has shown some promise at low temperatures is $\beta$-CuAgSe. The $\beta$ is used to distinguish it from the high temperature (above 470 K) $\alpha$-CuAgSe phase [22] whose properties more closely resemble the material's better-known cousins $Cu_2Se$ [23] and $Ag_2Se$ [24]. The low temperature $\beta$-CuAgSe phase is the subject of our discussion so the reader should assume that any mention of CuAgSe during the remainder of this report is a reference to this particular phase. $\beta$-CuAgSe possesses a highly disordered layered structure consisting of alternating Ag and CuSe layers. It is worth noting that there is some dispute over whether the correct classification is tetragonal or orthorhombic, but the difference is very subtle and is likely of little relevance to our discussion of the material's thermoelectric properties [25-26]. The first major investigation of the material's low temperature properties was conducted by Ishiwata *et al.* and published in 2013 [22]. Their report consisted of both electronic structure calculations as well as a series of experiments that revealed several notable and somewhat unexpected facts about this fascinating material. One such revelation was that the material exhibits reasonably high Seebeck magnitudes (-30 $\mu$V/K at 100 K and -100 $\mu$V/K at 300 K). The authors attribute this to asymmetries in carrier populations (as revealed by band structure calculations) as well as transport characteristics (as revealed by magnetoresistance measurements). Hints of this possibility can be found in the band structure calculations which have $E_F$ located in a light electron band centered at the $\Gamma$ point and a much broader hole band located right below $E_F$ in between the $\Gamma$ and M points. And while these S values may be much lower than those exhibited by conventional semiconductor thermoelectrics, CuAgSe's incredibly low electrical resistivity and thermal conductivity (the features that earned it the identifier "high mobility phonon-glass semimetal") compensate for this and overall lead to reasonably high low-temperature ZT (0.02 at 100 K).

What really makes CuAgSe appear so promising as a thermoelectric; however, is what Ishiwata *et al.* found to occur when the material is doped with Ni. They found that their $Cu_{0.9}Ni_{0.1}AgSe$ sample exhibited an enhancement in the low temperature mobility (90,000 $cm^2V^{-1}s^{-1}$ at 10 K as opposed to 20,000 $cm^2V^{-1}s^{-1}$ in the un-doped sample) as well as in the Seebeck below 200 K. These changes then caused ZT to increase by a factor of 5 to 0.10 at 100 K, one of the highest values ever reported at that temperature. The reasons for these enhancements are not obvious and the dramatic increase in mobility is especially unusual given that the disorder induced by chemical doping usually decreases carrier mobility. The authors suggest that this may be a special case where doping-induced disorder changes the material's properties in a highly unusual way. In order to further investigate and hopefully obtain a more precise understanding of the nature of the effects of chemical substitution on the thermoelectric properties of this fascinating material, we have performed similar experiments on CuAgSe samples doped with Co and Cr.

## 2. Experimental

Polycrystalline ingot samples were obtained using the following method, which is identical to the one used by Ishiwata *et al.* [22]. Stoichiometric amounts of high purity (99.999%) elemental copper, silver, selenium, cobalt and chromium were first sealed inside an evacuated quartz ampule and loaded into a vertical furnace. The mixture was first heated to 673 K at 20 K per hour, then to 1073 K at 50 K per hour, kept at this temperature for 24 hours, and then cooled to room temperature at a rate of 100 K per hour. The ingot is then ball-milled for five hours, after which the melting procedure is repeated on the resulting powder. Specific compositions used in this report include $Cu_{1-x}Co_xAgSe$ (x=0.02, 0.05, 0.10) and $Cu_{1-x}Cr_xAgSe$ (x=0.02, 0.05). XRD patterns for resulting samples are shown in Fig. 1 and 2. Patterns for the doped samples generally match that of the un-doped sample with the exception of a few very small additional peaks at 30°, 33°, and 35° in the 10% Co sample, 28° in the 2% Cr sample, and 30°, 38° and 46° in the 5% Cr sample. It is unclear whether these peaks are due to the doping or the presence of secondary phases, but if the latter is the cause, the low intensities imply that the quantities are minimal (most likely less than 5%).

Temperature and magnetic field dependent electrical resistivity, thermal conductivity, and Seebeck coefficient measurements were then performed using a Quantum Design Physical Property Measurement System (PPMS).

## 3. Results and Discussion

Fig. 3 shows the electrical resistivity as a function of temperature. Similar to what was observed for the Ni-doped case in reference 11, Co and Cr doping generally lead to a significant decrease in resistivity. However, there do exist a few notable features such as an increased resistivity in the doped samples relative to the un-doped one at low temperatures (which was not the case for Ni) and a particularly high resistivity in the 5% Cr sample. Possible explanations for these effects as well as their relation to our other results will be discussed later.

Fig. 4 shows the thermal conductivity as a function of temperature. Like the Ni-doped CuAgSe $\kappa$ data in reference 22, doping generally leads to an increase in thermal conductivity. And similar to what was observed in our $\rho_{xx}(T)$ data, the exception to this is at very low temperatures where $\kappa$ is higher in the un-doped sample. This is to be expected because thermal transport in CuAgSe is largely electron dominated and in such systems, changes in thermal conductivity should mirror changes in electrical conductivity.

Fig. 5-8 show the longitudinal ($\rho_{xx}$) and hall ($\rho_{xy}$) magnetoresistance data at both 2 K and 300 K. One feature worth highlighting is the highly nonlinear curve for the un-doped sample at 2 K. Ishiwata *et al.* observed this as well and argued that it was indicative of there being two types of carriers (electrons and holes) with different lifetimes. Interestingly, this is not observed for any of the other samples, which may imply that this asymmetry is either not present or that evidence of it in this particular measurement is smeared out by the local magnetic moments associated with the dopant atoms. This data can also be analyzed in a way that allows us to estimate mobility and carrier density in each of the samples at a specific temperature. For consistency, we employed the same method used by Ishiwata *et al.* in their report. This is done by first using the following equations to calculate conductivities $\sigma_{xx}$ and $\sigma_{xy}$ from the resistivities obtained in our measurements:

$$\sigma_{xx} = \frac{\rho_{xx}}{\rho_{xx}^2 + \rho_{xy}^2}$$

$$\sigma_{xy} = -\frac{\rho_{xy}}{\rho_{xx}^2 + \rho_{xy}^2}$$

These quantities can then be fit to the equations associated with a two-carrier model:

$$\sigma_{xx} = \frac{ne\mu}{1 + (\mu B)^2} + C_{xx}$$

$$\sigma_{xy} = n_H e \mu_H^2 B \left[ \frac{1}{1 + (\mu_H B)^2} + C_{xy} \right]$$

Here, n ($n_H$) is the longitudinal (Hall) carrier concentration, $\mu$ ($\mu_H$) is the longitudinal (Hall) mobility, e is the charge of the electron, B is the applied magnetic field, and $C_{xx}$ and $C_{xy}$ are low mobility components. The results of this analysis are shown in the tables below:

Table 1: Fitting parameters for Co-doped CuAgSe samples at 2 K

| Parameter | CuAgSe | $Cu_{0.98}Co_{0.02}AgSe$ | $Cu_{0.95}Co_{0.05}AgSe$ | $Cu_{0.90}Co_{0.10}AgSe$ |
|---|---|---|---|---|
| $\mu$ (cm²V⁻¹s⁻¹) | 34770 | 12900 | 10100 | 13300 |
| $\mu_H$ (cm²V⁻¹s⁻¹) | 36700 | 13300 | 10500 | 14100 |
| $n$ (m⁻³) | $8.34\times10^{24}$ | $9.45\times10^{24}$ | $1.17\times10^{25}$ | $9.45\times10^{24}$ |
| $n_H$ (m⁻³) | $8.04\times10^{24}$ | $8.86\times10^{24}$ | $1.07\times10^{25}$ | $8.15\times10^{24}$ |
| $C_{xx}$ (m-Ω⁻¹) | 133460.9 | 64299.4 | 123662.4 | 76476.6 |
| $C_{xy}$ | $-5.38\times10^{-5}$ | $3.4\times10^{-4}$ | 0.00212 | 0.00142 |

Table 2: Fitting parameters for Co-doped CuAgSe samples at 300 K

| Parameter | CuAgSe | $Cu_{0.98}Co_{0.02}AgSe$ | $Cu_{0.95}Co_{0.05}AgSe$ | $Cu_{0.90}Co_{0.10}AgSe$ |
|---|---|---|---|---|
| $\mu$ (cm²V⁻¹s⁻¹) | 2930 | 3100 | 3190 | 3030 |
| $\mu_H$ (cm²V⁻¹s⁻¹) | 2980 | 3190 | 3370 | 3130 |
| $n$ (m⁻³) | $3.13\times10^{24}$ | $7.4\times10^{24}$ | $6.88\times10^{24}$ | $7.08\times10^{24}$ |
| $n_H$ (m⁻³) | $3.08\times10^{24}$ | $6.8\times10^{24}$ | $5.18\times10^{24}$ | $6.4\times10^{24}$ |
| $C_{xx}$ (m-Ω⁻¹) | 10300 | 23200 | 57412 | 30797 |
| $C_{xy}$ | 0.00426 | 0.011 | 0.01693 | 0.0125 |

Table 3: Fitting parameters for Cr-doped CuAgSe samples at 2 K

| Parameter | CuAgSe | $Cu_{0.98}Cr_{0.02}AgSe$ | $Cu_{0.95}Cr_{0.05}AgSe$ |
|---|---|---|---|
| $\mu$ (cm²V⁻¹s⁻¹) | 34770 | 13800 | 5880 |
| $\mu_H$ (cm²V⁻¹s⁻¹) | 36700 | 16100 | 6940 |
| $n$ (m⁻³) | $8.34\times10^{24}$ | $1.05\times10^{25}$ | $4.05\times10^{24}$ |
| $n_H$ (m⁻³) | $8.04\times10^{24}$ | $8.04\times10^{24}$ | $6.2\times10^{23}$ |
| $C_{xx}$ (m-Ω⁻¹) | 133460.9 | 94307.4 | 100938.3 |
| $C_{xy}$ | $-5.38\times10^{-5}$ | $6.45\times10^{-4}$ | 0.0257 |

Table 4: Fitting parameters for Cr-doped CuAgSe samples at 300 K

| Parameter | CuAgSe | $Cu_{0.98}Cr_{0.02}AgSe$ | $Cu_{0.95}Cr_{0.05}AgSe$ |
|---|---|---|---|
| $\mu$ (cm$^2$V$^{-1}$s$^{-1}$) | 2930 | 2580 | 1740 |
| $\mu_H$ (cm$^2$V$^{-1}$s$^{-1}$) | 2980 | 2690 | 1800 |
| $n$ (m$^{-3}$) | 3.13×10$^{24}$ | 7.52×10$^{24}$ | 3.76×10$^{24}$ |
| $n_H$ (m$^{-3}$) | 3.08×10$^{24}$ | 6.79×10$^{24}$ | 1.8×10$^{24}$ |
| $C_{xx}$ (m-$\Omega^{-1}$) | 10300 | 19472 | 65415 |
| $C_{xy}$ | 0.00426 | 0.00433 | 0.0148 |

This analysis helps to explain some of the notable features in the transport data. For example, the main difference between the Ni-doped CuAgSe data in reference 22 and our Co and Cr-doped CuAgSe data is the increase in electrical resistivity that was observed in our samples at low temperatures relative to the un-doped one. The values given in the table above imply that this difference can be attributed to the effects the different dopants have on mobility. Ishiwata *et al.* found that Ni doping actually increased the mobility at 10 K from 20,000 Vm$^{-1}$s$^{-1}$ to 90,000 Vm$^{-1}$s$^{-1}$. Our data; however, reveals that Co and Cr-doping both decrease the low temperature mobility substantially. This decrease is particularly significant in the 5% Cr sample, which explains why its resistivity is so much higher than the other samples in this report. The fact that these differences are only exhibited at low temperatures implies that their origin is likely in characteristics whose effects are more significant at those low temperatures. One possible explanation is the disorder-based mechanism given by Ishiwata *et al.* They suggest that the increased mobility caused by Ni-doping is due to a somehow unique character of the substitution-induced disorder in Ni-doped CuAgSe. If this is in fact the case, it may just be that instead of producing this unusual form of disorder, Co and Cr-doping just lead to a more conventional form and cause the mobility to decrease, as is typically the case.

Other possible explanations that are more precise include the exact nature and location of the defect states, possible magnetic ordering, as well as various combinations of each of these effects. More experiments will of course be needed to better study these other properties.

It is also worth noting that the effects at high temperatures are somewhat different. For the Cr-doped samples, a decrease in mobility is again observed. However, for the Co-doped ones, the mobility actually increases slightly, as it was previously observed to do with Ni-doping [11]. It is still unclear what exactly causes this increase, but these results imply that whatever physics caused this to occur for Ni may be at play with Co as well.

Fig. 9 shows Seebeck as a function of temperature in both a 0 T and a 9 T magnetic field. This plot is quite complex, but the following explanation should help the reader to better understand its underlying story.

We begin with the un-doped CuAgSe sample without the magnetic field (black squares). Starting at 0 K (where S is approximately 0), as the temperature is increased, the magnitude of S increases (meaning more and more negative) as well. The shape of the curve is fairly linear until about 50 K where the curve kinks downward relative to its initial trend. For the Co and Cr-doped sample curves (colored squares), all S curves coincide with that of the un-doped sample until 50 K where they all kink upward instead of downward like the un-doped one did. It should be noted that while the degree to which this occurs seems to increase with doping, it does appear to saturate at higher concentrations. The saturation seems to imply that the origin of this effect is more likely related to disorder and/or magnetic effects as opposed to changes in carrier density for which one would typically expect a more linear change with doping. Interestingly, the S curve for Ishiwata *et al.*'s $Cu_{0.9}Ni_{0.1}AgSe$ sample (gray line) is completely different [22]. Instead of following the same linear trend as the curve for the un-doped sample below 50 K, the curve for the Ni-doped one kinks downward around 0 K and remains below the un-doped sample curve until approximately 220 K where the two intersect. If the differences in the S curves are in fact related to disorder, these results would further the case that the disorder in Ni-doped CuAgSe is somehow unique. It is again also possible that these changes may be rooted in magnetic and/or impurity state effects. Given that CuAgSe's thermoelectric performance is thought to be driven partially by scattering physics (i. e. asymmetries in carrier lifetimes), one additional possibility that may be worth exploring is a mechanism similar to the one presented in reference 27 where Seebeck is driven by mobility gradients in addition to DOS gradients [27]. Nernst coefficient measurements such as those suggested in that report should be informative for investigating this possibility.

We also performed these same measurements in a 9 T magnetic field (stars) and obtained some rather fascinating results. A very similar response is seen for all of our Co and Cr-doped samples. Instead of diverging from the linear trend observed at low temperatures, the 9 T doped sample curves roughly maintain their linearity up to 300 K where our measurements stop. This implies that whatever physics causes this "kinking" feature at 50 K in the doped samples is somehow suppressed by the application of a magnetic field. Unfortunately, no 9 T Seebeck data is available for Ni-doped CuAgSe. This effect; however, is not observed for the un-doped sample (black stars). For this sample, the 0 T and 9 T curves are completely different and at very low temperatures, the magnetic field appears to have the effect of actually causing the Seebeck coefficient to change sign (negative to positive). It is somewhat rare for Seebeck to respond significantly at all to a magnetic field, so a change in sign is particularly interesting. This; however, is not the first reported incidence of this effect. A similar phenomenon was recently observed in LuPtBi, another high mobility semimetal, where the unusual Seebeck response was attributed to asymmetric temperature and field-dependent responses of the electrons and holes [28]. The same explanation appears to describe the Seebeck data for the un-doped CuAgSe and is consistent with the nonlinear $\rho_{xy}(B)$ curve found for the same sample. This picture; however, does not explain the magnetic response of the Seebeck in the doped samples where the $\rho_{xy}(B)$ curves are highly linear and the response of Seebeck to a magnetic field is less significant. One possible explanation is that the local moments associated with the dopant atoms simply smear out these additional features, but a more detailed understanding of the nature of the magnetism in the Co- and Cr-doped samples will be needed before further conclusions can be made.

## 4. Conclusions

In summary, we have synthesized and measured the low temperature thermoelectric properties of samples of CuAgSe doped with varying amounts of Co and Cr. Our results reveal a significant sensitivity of these properties to doping (as was previously shown to be the case with Ni-doping) as well as a particular sensitivity to specific dopants. Also observed was an unusually strong response of the material's thermoelectric power to an applied magnetic field, an effect that is likely due to the asymmetric two-carrier nature of electronic transport in CuAgSe. Possible explanations for the doping-induced changes include effects related to disorder, defect states, magnetic ordering, and scattering phenomena, but further experiments will be needed in order to better probe the exact mechanism(s) by which they result. The remarkable tunability of CuAgSe's thermoelectric transport properties revealed by this investigation furthers the case that high mobility phonon glass semimetals are promising candidates for low temperature thermoelectric applications.

**Acknowledgements**

We gratefully acknowledge funding for this work by the Department of Defense, United States Air Force Office of Scientific Research, Multi-University Research Initiative (MURI) Program under Contract # FA9550-10-1-0533.

Figures

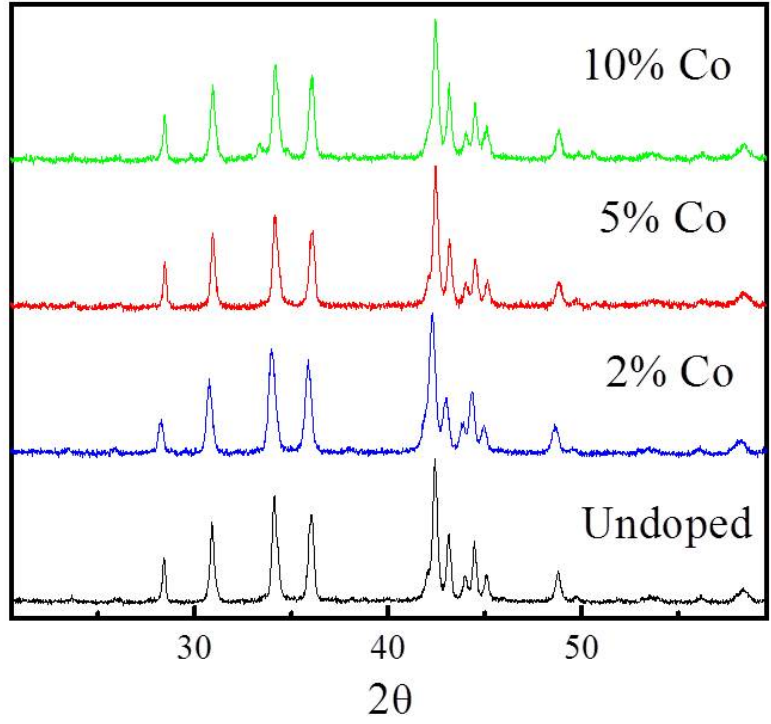

Fig. 1: XRD Patterns for Co-doped samples.

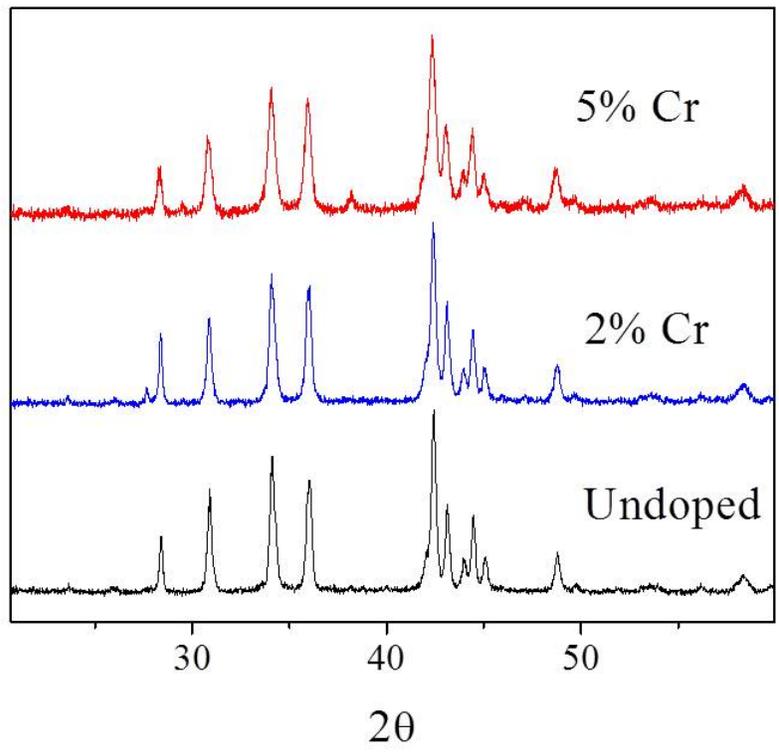

Fig. 2: XRD Patterns for Cr-doped samples.

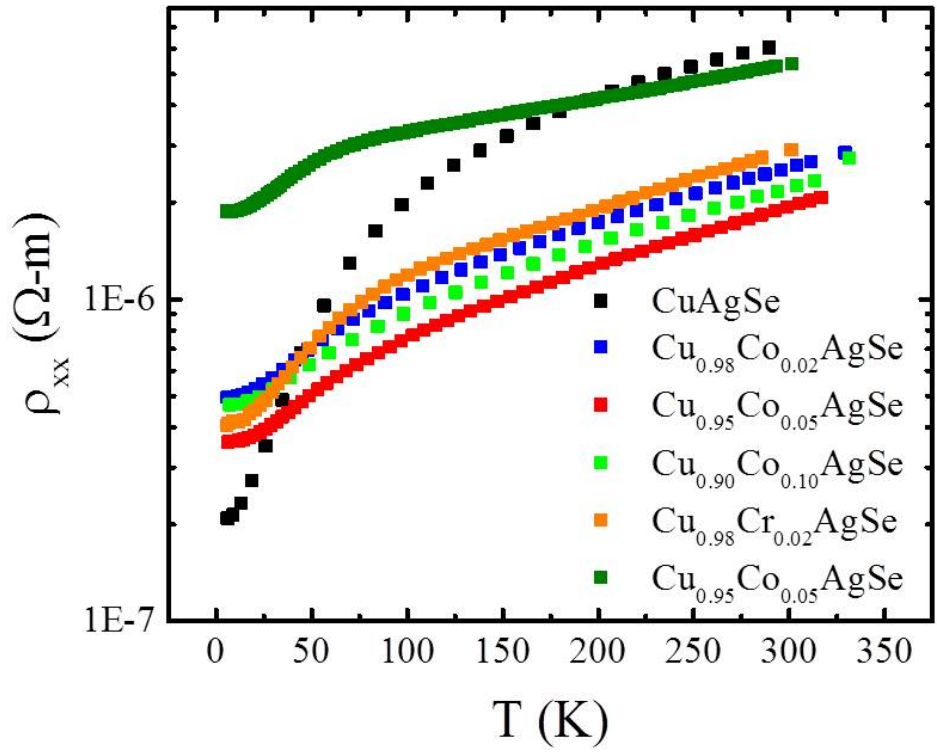

Fig. 3: Electrical resistivity of all samples as a function of temperature.

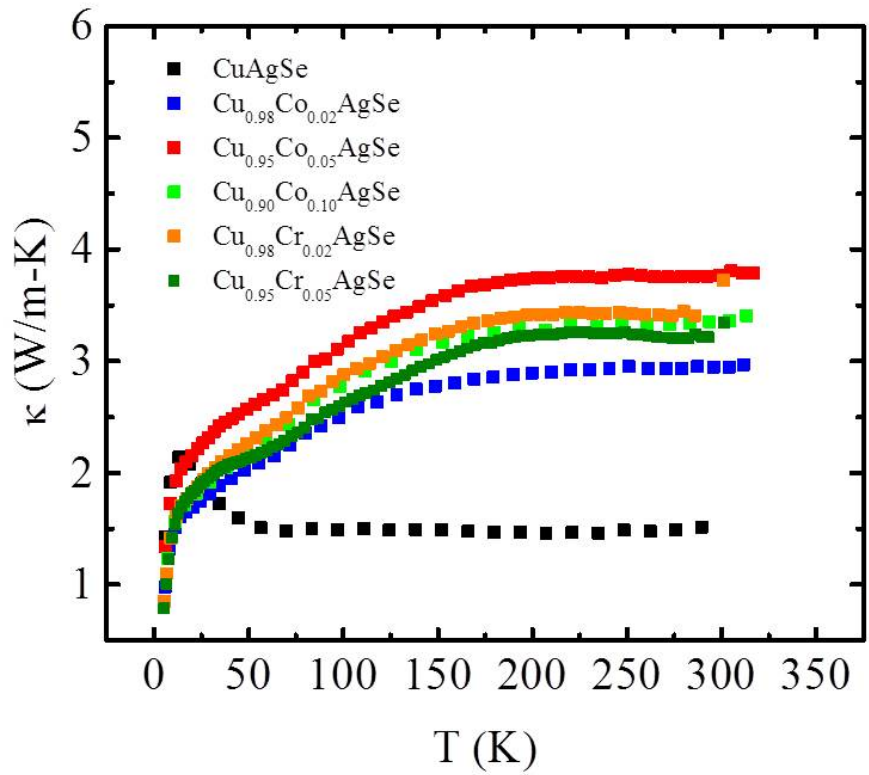

Fig. 4: Thermal Conductivity of all samples as a function of temperature.

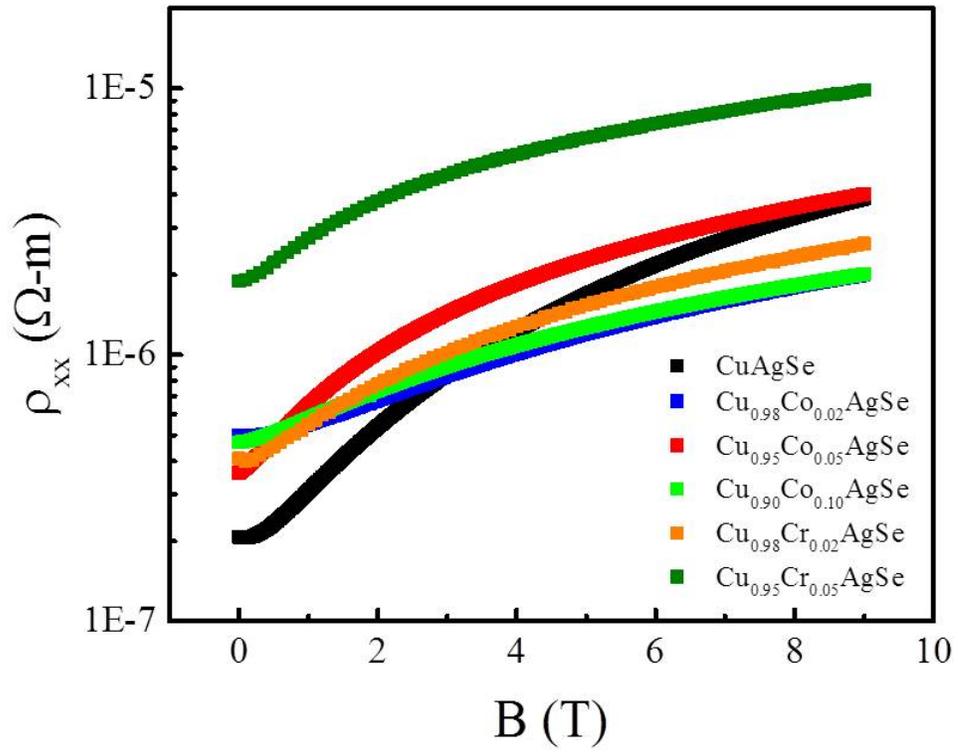

Fig. 5: Longitudinal resistivity of all samples as a function of magnetic field at 2 K.

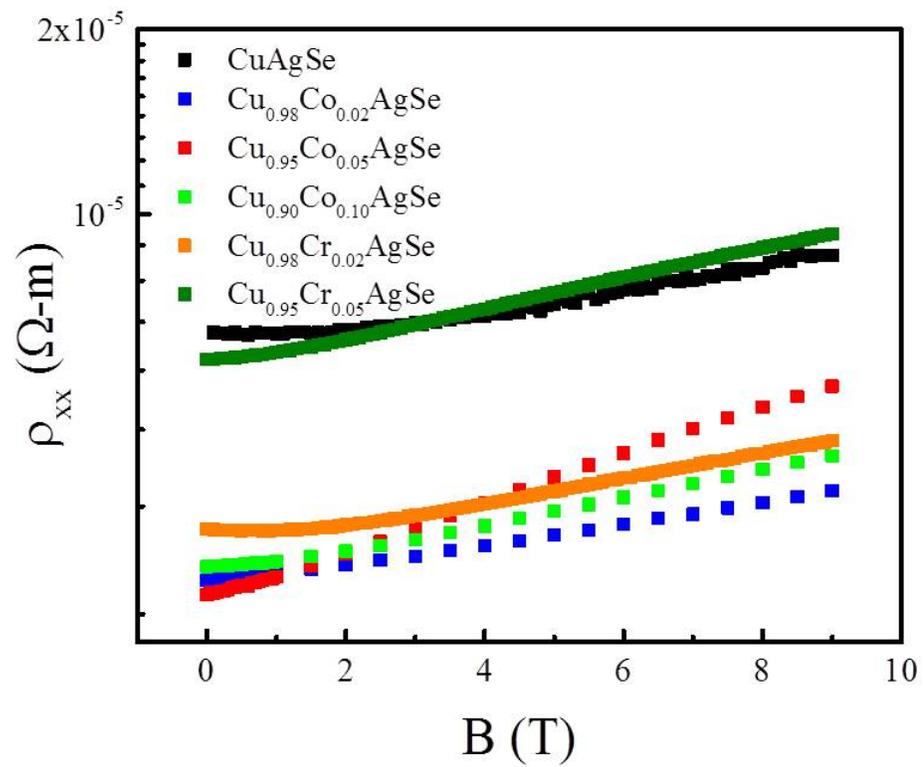

Fig. 6: Longitudinal resistivity of all samples as a function of magnetic field at 300 K.

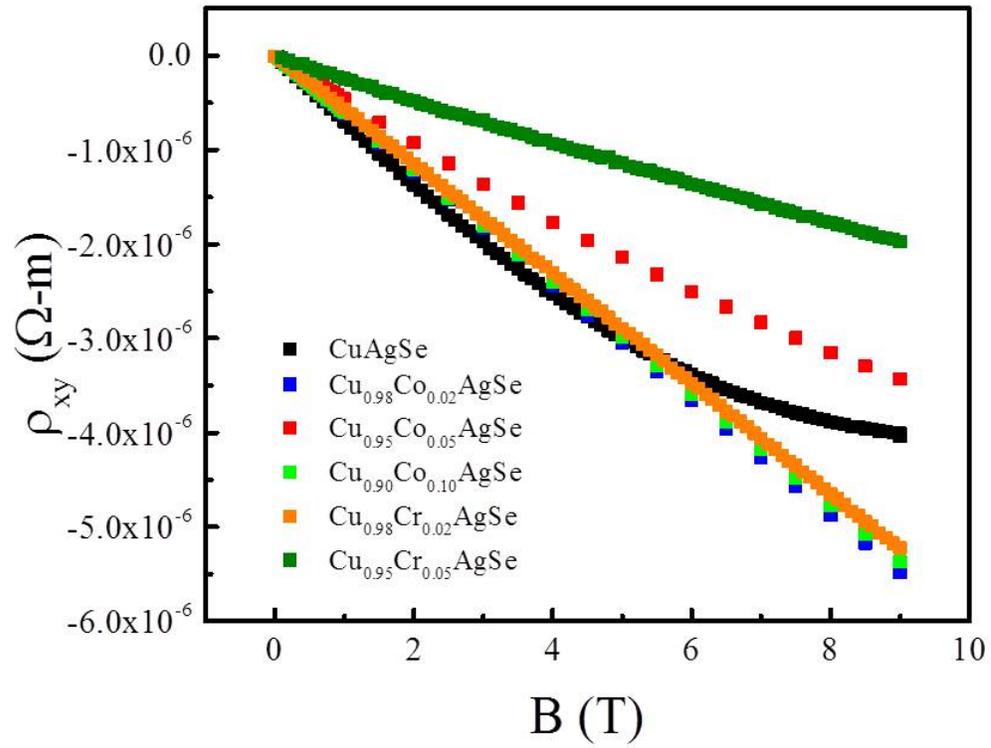

Fig. 7: Hall resistivity of all samples as a function of magnetic field at 2 K.

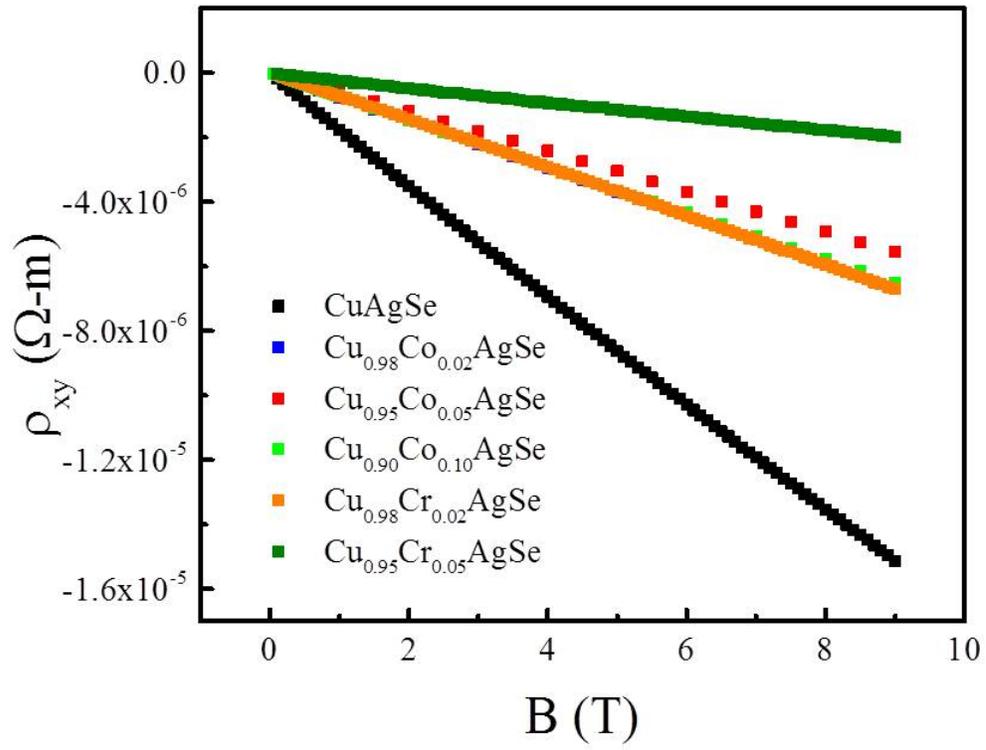

Fig. 8: Hall resistivity of all samples as a function of magnetic field at 300 K.

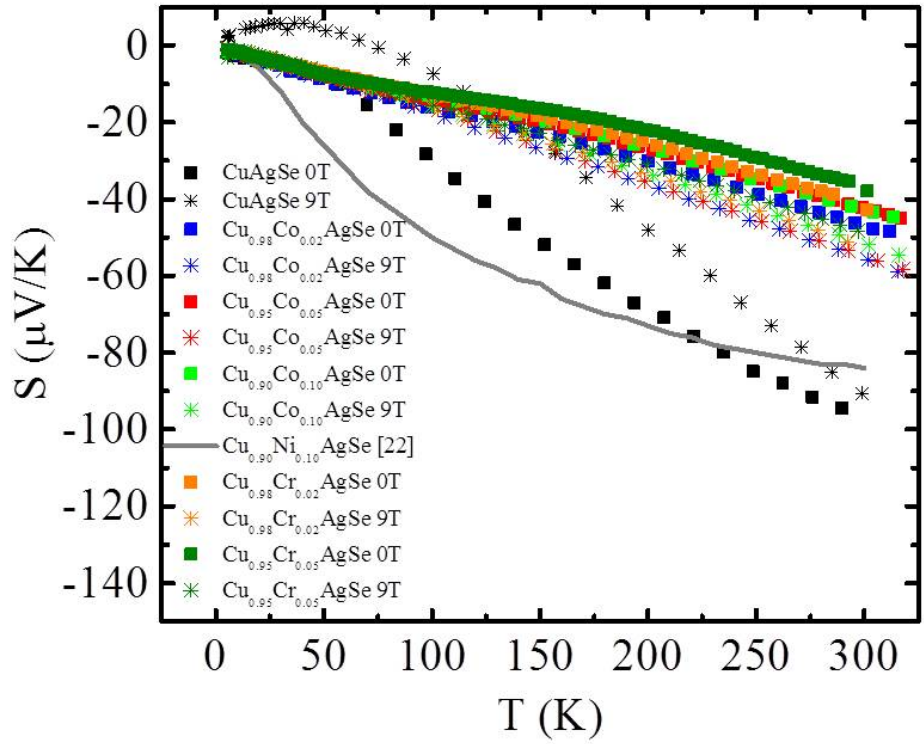

Fig. 9: Seebeck coefficient for all samples in both 0 and 9 T magnetic field as a function of temperature. Also shown is the Ni-doped CuAgSe S (T) data (gray line) from reference 22.